\documentclass{article}
\usepackage{amssymb,latexsym}
\usepackage{amsmath}

\begin{document}
\title{%
        Schr\"odinger equation for joint\\
        bidirectional motion in time}
\author{G. E. Hahne\thanks{email:  ghahne@mail.arc.nasa.gov}\\
        NASA, Ames Research Center\\
        Moffett Field, California, 94035 USA}
\maketitle

\centerline{PACS Numbers:  03.65.Nk, 03.70.+k, 11.10.Ef, 98.80.Hw}

\begin{abstract}

    The conventional time-dependent Schr\"{o}dinger equation
describes only unidirectional
time evolution of the state of a physical system, i.e., forward, or,
less commonly, backward.  This paper proposes a 
generalized quantum dynamics
for the description of joint, and interactive, 
forward and backward time evolution within
a physical system.  
The principal mathematical  assumption for bidirectional
evolution in general is that the space of states
should be taken to be not merely a Hilbert space, but a more
restricted entity
known as a Kre\u{\i}n space, which is a complex Hilbert space
with a Hermitean 
operator that has eigenvalues $+1$ and $-1$ only, and that therefore
gives rise to an indefinite metric.  The vector subspaces of states with
positive or negative norm with respect to the indefinite metric
will---for open channels---be construed to be states in
forward or, respectively, backward evolution along the time axis.
The quantum dynamics is generated by a pseudo-Hermitean
Hamiltonian operator and conserves inner products with respect
to the indefinite metric.
Input and output states are defined in physically plausible ways 
such that the output comprises both reflected and transmitted states
from a zone of interaction in time;
a unitary transformation between input and output states
is obtained from the pseudounitary transformation
between the initial and final states.
Three applications are studied:
(1) a formal theory of collisions in terms of
perturbation theory; (2) a relativistically invariant
quantum field theory for a system that kinematically comprises the direct
sum of two quantized real scalar fields, such that
one subfield evolves forward and the
other backward in time, 
and such that there is dynamical coupling
between the subfields;  (3) an argument that
in the latter field theory, the dynamics predicts that in
a range of values of the coupling constants, the
expectation value of the vacuum energy of the
universe is forced to be zero to high accuracy.
[Added in arXiv version:  It is also speculated that the ideas 
presented contain a kernel 
of explanation for the existence of a negative average energy density
in the cosmos.]
\end{abstract}

\section{Introduction} \label{S:sec1}

    The usual time-dependent Schr\"{o}dinger equation is
\begin{equation}\label{E1:eq1}
\left[\frac{\hbar}{i}\frac{\partial}{\partial t}\,+\ H\right]\Phi(t)\ =\ 0,
\end{equation}
where $\Phi(t)$ is a time-evolving state vector in a Hilbert space, and
$H$ is a Hermitean Hamiltonian operator.  This equation has the
property that it describes just the unidirectional evolution of the
state of a physical system from one time to another time that can be later
or earlier than the first.  Our principal
objective herein is to construct a more general
form of quantum mechanics that can describe a physical system
in which part of the system evolves forward in time, while the
remaining part evolves backward in time, and such that the two parts
can interact.  
     
     The argument proceeds from the
observation that such a formalism can be inferred from the
quantum mechanics of two known physical systems:
the first is the description, by a time-independent
Schr\"{o}dinger equation, of the evolution of a system along a space-like
reaction coordinate, and the second is the complex Klein-Gordon equation
for the motion of a spinless particle in the presence of a
fixed, transient vector potential field.
We shall not present the theory associated with these cases in detail,
but sketch the ideas in the following two paragraphs.

     For the evolution of a steady-state physical system along
a space-like reaction coordinate, we cite as an example the evolution
of the reversible, collinear chemical reaction 
$A+BC\leftrightarrow AB+C$ in the center-of-mass system---see Marcus 
\cite{R:Marcus1}, Light \cite{R:Light1}, 
Baer \cite{R:Baer1}, and Miller \cite{R:Miller1}---or,
more simply, reflection and transmission of a beam of structureless
particles from a potential barrier in one dimension.  
The second-order Schr\"{o}dinger 
equation can be recast \cite{R:Hahne1} as a coupled system
of ordinary first-order equations, where the wave function is expanded
in a set of vibrational states of the transverse coordinate.
An indefinite metric matrix is derived from Wronskians, such that
waves travelling forward along the reaction coordinate have
positive norms, and waves travelling backward along the reaction
coordinate have negative norms.  The dynamics is governed by a Hamiltonian
that is pseudo-Hermitean with respect to the metric, and hence conserves
inner products with respect to the metric.  The input comprises
travelling waves (i.e., open channels)
converging on the reaction zone, and the output comprises waves
diverging from the reaction zone.  A unitary $S$-matrix
transforming input into output can be assembled from 
reflection and transmission matrices pertaining to open channels.

    In the case of the Klein-Gordon equation, a
Schr\"{o}dinger-equation-like formalism has been derived by Feshbach
and Villars \cite{R:Fesh1}, Eq.\ (2.15), et seq.
The Hamiltonian proves to be pseudo-Hermitean
with respect to an indefinite metric.  
The input comprises positive energy (and positive norm)
states at large negative times, and negative energy (and negative
norm) states at large positive times;  the output comprises
negative norm states at large negative times, and positive
norm states at large positive times.
It is straightforward, using the
formalism developed by Bjorken and Drell \cite{R:BjD1}, Eqs (9.6)
and (9.20), 
to verify that a unitary $S$-matrix mapping input into output
can be constructed from reflection and transmission coefficients.

     A considerable selection of books has been published
that is concerned with the physics and
metaphysics of time, irreversibility,
time's arrow, and so on.  A representative list
comprises Reichenbach \cite{R:Reichenbach1},  
Landsberg \cite{R:Landsberg1}, Davies \cite{R:Davies1,R:Davies2}, 
Zeh \cite{R:Zeh1}, Schulman \cite{R:Schulman1}, Price \cite{R:Price1},
Novikov \cite{R:Novikov1}, and Penrose \cite{R:Penrose1,R:Penrose2}.
The book by Zeh \cite {R:Zeh1} has a long list of references
on the subject of its title, and much quantitative discussion;  
Zeh has put a preliminary version of a fourth edition of his book
online (www.time-direction.de).

     A line of investigation related to the present one
was initiated by Schr\"odinger \cite{R:Schroedinger1},
as elaborated by Aebi \cite{R:Aebi1,R:Aebi2},
and other works referenced therein.
The generic idea of this line is to consider the evolution
of diffusion or quantum processes, for which partial
information on the state of the system is given at each of two
finite times, and to infer the likeliest state of the system
at intermediate times.  Aharonov, et al. \cite{R:Aharonov1},
and Reznik, et al. 
\cite{R:Reznik1} constructed a time-symmetric quantum
mechanics that utilizes information about the
state of a system at both ends of a time interval to infer
the expected results of measurements at an intermediate time.
These investigations did not attempt to generalize quantum mechanics
as is done here, but recast the existing physical laws
in an alternate form.  
Perhaps the closest predecessor theory to that
presented herein is the discussion/analysis of the 
problem of two-point boundary conditions in quantum
mechanics in Schulman (\cite{R:Schulman1}, Ch. 5.3).
Schulman's work is discussed in Zeh
(\cite{R:Zeh1}, Ch.\ 5.3).  
In particular, Schulman (\cite{R:Schulman1}, p.\ 184)
introduces ``subspace boundary value problems'' as a
category of two-time boundary conditions;  nevertheless, Schulman's
quantum dynamics uses a Hermitean Hamiltonian, and 
correspondingly does
not introduce an indefinite metric, so that his proposed theory
does not conserve probability in the sense that 
will be done here.
Schulman \cite{R:Schulman2}, plus a directed comment by
Casati, et al.\ \cite{R:Casati1} and reply by Schulman
\cite{R:Schulman3}, dealt with a classical mechanics
construction of opposite thermodynamic arrows of time.

     Cramer \cite{R:Cramer1} has developed a
``transactional'' interpretation
of quantum mechanics that involves the presence of advanced
as well as retarded, interactions that are invoked to
relieve some of the counterintuitive nonlocality involved
in the collapse of the wave function.  Cramer, however, does not
introduce a generalized dynamics associated with the 
transactional interpretation,
and makes predictions that do not differ from those of standard
quantum mechanics (Ref.\ \cite{R:Cramer1}, Ch.\ III.B, last paragraph).
I infer also that Cramer presumes
that the strength of the interactions of the advanced waves
with ordinary matter are the same as, or roughly comparable to,
those of retarded waves.  
In the theory described below, interactions between the
forward- and backward-evolving subspaces are presumed 
on physical grounds to
be very small compared to, say, electromagnetic interactions
within each subspace.

   A subject that is employed in the mathematics used
herein is the study of infinite-dimensional complex vector spaces
that are endowed with a nondegenerate sesquilinear inner product that
gives rise to an indefinite metric.  In quantum field theory,
this subject was first studied by Dirac \cite{R:Dirac1},
and in mathematics by Pontrjagin \cite{R:Pontrjagin1}.  The
former area was a subject of 
interest in the `40's to the early `60's, as reviewed
in Nagy \cite{R:Nagy1};  the latter subject is still an
area of mathematical interest---see Azizov, et al.
\cite{R:Azizov2}.

     The remainder of this article is organized as follows.    
In Sec.\ \ref{S:sec2} we formulate a quantum dynamics,
in the form of a Schr\"{o}dinger equation and some rules
for interpreting the associated mathematics, which can treat
physical systems in which joint, and interactive, motion or
evolution in both directions in time can occur.  Sec.\ \ref{S:sec3} 
derives a formal theory of scattering, i.e., transition operators and
S-matrices, for collision processes with a time-independent
Hamiltonian governing the dynamics.  Sec.\ \ref{S:sec4}
presents the basics of a physical system comprising the
direct sum of two interacting quantized real scalar fields;  the theory is
shown to be relativistically invariant, and perturbation
theory is applied to a case of two-body collisions.
Sec.\ \ref{S:sec5} concludes the paper 
with a discussion of some of the ideas presented herein, 
and with a quantitative argument to the effect that
in a suitable range of parameter values of the field theory
of Sec.\ \ref{S:sec4}, the expectation value of the vacuum energy
of the universe necessarily vanishes to high accuracy.
The Appendix shows how to obtain transition rates from
transition operators.

      We emphasize that the statement given herein of a Schr\"{o}dinger
equation to describe bidirectional motion in time is incomplete:
important, but derivative, theoretical aspects, such as 
a manifestly covariant perturbation scheme for the quantum field theory
of Sec.\ \ref{S:sec4}, and
modifications of quantum measurement theory, 
including an analysis of wave function ``collapse'', etc., 
remain to be worked out.

\section{Quantum mechanics of bidirectional motion}\label{S:sec2}

     In this section we shall propose a formalism
that accomplishes the paper's 
title objective.  The principal mathematical idea is
to introduce a state space with a nondegenerate inner
product that yields an indefinite metric,
and correspondingly, a pseudo-Hermitean Hamiltonian to govern the
dynamics.   The attendant physical interpretation will
posit that state space comprises the direct sum
of two orthogonal subspaces, such
that one has a positive definite 
norm and the other a negative definite norm;  for open channels,
these two subspaces will correspond to those states of motion of 
the system that evolve forward and backward in time, respectively.

     Some of the mathematical community
presently designates a state space
of the above type, with a suitable topology,
as a Kre\u{\i}n space (described, with references, in
the encyclopedia \cite{R:encyc1}, Vol.\ 5, p.\ 303)
named for the Ukrainian
mathematician M. G. Kre\u{\i}n---see Azizov, et al.\ \cite{R:Azizov1}
for a description of Kre\u{\i}n's work in this area, and
Azizov and Iokhvidov \cite{R:Azizov2} for the theory of Kre\u{\i}n spaces.
An earlier designation, Nevanlinna space 
(mentioned in Nagy \cite{R:Nagy1}, \S 1),
now applies to a different entity (Juneja, et al.\ \cite{R:Juneja1}).
The properties of matrices in finite-dimensional vector spaces
with an indefinite metric are discussed in Gohberg,
et al.\ \cite{R:Gohberg1}.
An alternate formulation of the latter class of spaces
has been called ``complex symplectic geometry''
(see Everitt, et al. \cite{R:Everitt1}), although this usage conflicts
with an earlier development (Chevalley \cite{R:Chev1}, p.\ 23, Definition 1), 
in that the
extension of symplectic geometry from the real coefficient field
to the complex field entails a
sesquilinear and, implicitly, a bilinear form in the respective definitions.
The mathematical physics community for the most
part seems to have used the designation
``space with an indefinite metric'', although the name ``Kre\u{\i}n
space'' sometimes appears (Mnat\-sa\-kan\-o\-va,
et al.\ \cite{R:MMSV1});  the designation
``pseudo-Hilbert space'' (Konisi, et al.\ \cite{R:Konisi1,R:Konisi2})
was used rarely. 
 
     Beginning with the work of Dirac \cite{R:Dirac1} and 
Pauli \cite{R:Pauli1}, a substantial body of work on quantum field
theory was done that dealt with state spaces with an indefinite metric, as
reviewed in Ref.\ \cite{R:Nagy1}.  There is little overlap between
this theoretical work and that presented below:  
(1) we shall not introduce anomalous commutators for the
creation and destruction operators associated with a quantum field;
(2)  we shall (in Sec.\ \ref{S:sec4}) deal with 
a field theory for which a complete quantum state 
is a vector in a space that is made up of the 
direct sum of the Fock spaces of two conventional
field theories;  (3) The S-matrix will be obtained, not by 
the mapping of the system's state at $t=-\infty$ into the state
at $t=+\infty$ as input into output, but as a mapping with a different
choice of input and output such that probability is conserved
and the S-matrix is unitary.

     More recent work on associated mathematical physics, such as
Mnat\-sa\-kan\-o\-va, et al. \cite{R:MMSV1}, will not be needed herein as
the nonlocal
input/output conditions in time suggest a different approach.

     We begin with a Hilbert space $\mathcal H$, with vectors denoted
say $\psi\in{\mathcal H}$, and a sesquilinear product 
$\langle.\,,.\rangle$
with the standard inner product (unit metric matrix) form, such that
\begin{subequations}\label{E4:eq1}
\begin{align} 
\langle\psi_1,\psi_2\rangle\ &=\ (\psi_1)^\dagger\psi_2,\label{E4:eq1a}\\
\langle\psi_1,\alpha\psi_2+\beta\psi_3\rangle
\ &=\
\alpha\langle\psi_1,\psi_2\rangle\,+\,\beta\langle\psi_1,\psi_3\rangle,
\label{E4:eq1b}\\
\langle\psi_1,\psi_2\rangle\ &=\ \langle\psi_2,\psi_1\rangle^\ast.
\label{E4:eq1c}
\end{align}
\end{subequations}
We postulate further that $\mathcal H$ is equivalent to the direct sum of 
exactly two subspaces ${\mathcal H}^{F}$ and ${\mathcal H}^B$
with corresponding Hermitean projection operators $P^F$ and $P^B$,
such that
\begin{subequations}\label{E4:eq2}
\begin{align}
(P^Y)^\dagger\ &=\ P^Y,\label{E4:eq2a}\\
P^F\,+\,P^B\ &=\ I,\label{E4:eq2b}\\
P^{Y}P^{Y'}\ &=\ P^Y\delta^{YY'},\label{E4:eq2c}\\
P^{F}{\mathcal H}\ &=\ {\mathcal H}^F\oplus
0^B,\label{E4:eq2d}\\
P^{B}{\mathcal H}\ &=\ 0^F\oplus{\mathcal H}^B,
\label{E4:eq2e}
\end{align}
\end{subequations}
where $I$ is the identity operator in $\mathcal H$, $Y$ and $Y'$
can each be $F$ or $B$,
and $0^Y$ is the zero subspace in $\mathcal{H}^Y$.  
We shall use $I^F$ and
$I^B$ as the identity operator in the respective subspace.  
We shall not distinguish between
$\mathcal H$ and the direct sum 
${\mathcal H}^F\oplus{\mathcal H}^B$;
accordingly, if we define for $Y=F,B$,
\begin{equation}\label{E4:eq3}
\psi^{Y}\ =\ (P^Y\psi)|_{\mathcal{H}^Y}\in\mathcal{H}^Y,
\end{equation}
we can describe $\psi$ in block column matrix form as
\begin{equation}\label{E4:eq4}
\psi\ =\ \left[\begin{matrix} \psi^F \\ \psi^B\end{matrix}
\right].
\end{equation}

    We now define an operator $\eta$ that engenders an indefinite metric:
\begin{equation}\label{E4:eq5}
\eta = P^F\,-\,P^B,
\end{equation}
and an associated inner product $(.\,;.)$ as
\begin{equation}\label{E4:eq6}
(\psi_1;\psi_2)\ =\ \langle\psi_1,\eta\psi_2\rangle.
\end{equation}
The $\eta$-adjoint $T^\ddag$ of an operator $T$ acting on $\mathcal H$
is defined as that unique operator that satisfies
\begin{equation}\label{E4:eq7}
(T^\ddag\psi_1;\psi_2)\ =\ (\psi_1;T\psi_2)
\end{equation}
for all $\psi_1,\psi_2\in{\mathcal H}$.
An operator $T$ will be called pseudounitary
if it preserves $\eta$-products, that is,
for all $\psi_1,\psi_2\in{\mathcal H}$ we have
\begin{equation}\label{E4:eq8}
(T\psi_1;T\psi_2)\ =\ (\psi_1;\psi_2)
\end{equation}
and pseudo-Hermitean if $T^\ddag=T$, that is,
\begin{equation}\label{E4:eq9}
(T\psi_1;\psi_2)\ =\ (\psi_1;T\psi_2).
\end{equation}

   If we revert to the block matrix form of Eq.\ \eqref{E4:eq4}
we infer that
\begin{equation}\label{E4:eq10}
(\psi_1;\psi_2)\ =\ \psi_1^\dagger\eta\psi_2
\ =\ (\psi_1^F)^\dagger\psi_2^F
\,-\,(\psi_1^B)^\dagger\psi_2^B.
\end{equation}
Also, if for an operator $T$ we define
\begin{equation}\label{E4:eq11}
T^{YY'}\ =\ (P^Y T P^{Y'})|_
{\rm{Hom}[\mathcal{H}^Y\leftarrow\mathcal{H}^{Y'}]},
\end{equation}
where $\rm{Hom}[\mathcal{H}^Y\leftarrow\mathcal{H}^{Y'}]$
is the set of complex-linear mappings
(i.e., homomorphisms) from $\mathcal{H}^{Y'}$
into $\mathcal{H}^Y$,
then we have, in block matrix notation,
\begin{equation}\label{E4:eq12}
T\ =\ \left[\begin{matrix}T^{FF} & T^{FB}          \\
T^{BF}          & T^{BB}\end{matrix}\right].
\end{equation}
If $T$ is pseudo-Hermitean we have
\begin{equation}\label{E4:eq13}
T\ =\ T^\ddag\ =\ \eta T^\dagger\eta
\ =\ \left[\begin{matrix}(T^{FF})^\dagger 
& -(T^{BF})^\dagger \\
-(T^{FB})^\dagger &
(T^{BB})^\dagger\end{matrix}\right].
\end{equation}

     If $T$ is pseudounitary we have the 
(we presume, both left and right) inverse $T^{-1}$
that satisfies
\begin{equation}\label{E4:eq14}
T^{-1}\ =\ T^\ddag,
\end{equation}
and, therefore,
\begin{subequations}\label{E4:eq15}
\begin{align}
(T^{FF})^\dagger(T^{FF})
\,-\,(T^{BF})^\dagger(T^{BF})
\ &=\ I^F,\label{E4:eq15a}\\
-(T^{FB})^\dagger(T^{FF})
\,+\,(T^{BB})^\dagger(T^{BF})
\ &=\ 0,\label{E4:eq15b}\\
-(T^{FB})^\dagger(T^{FB})
\,+\,(T^{BB})^\dagger(T^{BB})
\ &=\ I^B.\label{E4:eq15c}
\end{align}
\end{subequations} 
Let it be given that $T$ is pseudounitary and that 
$T^{BB}$ has an inverse
$(T^{BB})^{\iota}$ within ${\mathcal H}^B$, in that
\begin{equation}\label{E4:eq16}
(T^{BB})^\iota\, T^{BB}\ =\ I^B\ =\ 
T^{BB}(T^{BB})^\iota;
\end{equation}
then the block operator-matrix $\tilde U(T)$, defined as
\begin{equation}\label{E4:eq17}
\tilde U(T)\ =\ \left[\begin{matrix} 
T^{FF}-T^{FB}(T^{BB})^\iota T^{BF} &
T^{FB}(T^{BB})^\iota \\
-(T^{BB})^\iota T^{BF} &
(T^{BB})^\iota\end{matrix}\right],
\end{equation}
can, with the aid of Eq.\ \eqref{E4:eq15}, be proved to be unitary
on the left, and similarly for right unitarity.
A more complicated procedure is needed to extract a unitary
$S$-matrix when asymptotic closed channels are present---see 
Sec.\ \ref{S:sec3}.

     A time-dependent vector $\psi(t)\in\mathcal H$ that is an eigenvector of
$\eta$ with eigenvalue $+1$ (resp., $-1$) will,
asymptotic closed channels excepted, be considered to evolve
forward (resp., backward) in time.  
The expectation value $(\psi(t);\psi(t))$ 
of a general state $\psi(t)$ will be construed
as the integrated probability current crossing the complete space-like
surface time=$t$.   The operator $\eta$ is therefore a kind
of velocity operator, describable as the derivative of dynamical causation
time with respect to kinematical time, and can take only the values
$+1$ and $-1$.  This interpretation therefore addresses the
question of the velocity of objective flow of time
posed in Price (\cite{R:Price1}, p. 13).

     We proceed from kinematics to a theory of quantum dynamics.  
Let $P^F$ and $P^B$ be time-independent, let $H(t)$
be a Hamiltonian that is pseudo-Hermitean at each instant,
and let $\Phi(t)\in {\mathcal H}$ be a kinematically allowable family
of state vectors, described parametically by dependence on the time.
The time evolution of a dynamically allowable family of
quantum states $\Phi(t)$ is governed by the Schr{\"o}dinger equation
\begin{equation}\label{E4:eq18}
i\frac{d}{dt}\Phi(t)\ =\ H(t)\Phi(t).
\end{equation}
When both $\Phi_1(t)$ and $\Phi_2(t)$ are solutions of Eq.\
\eqref{E4:eq18}, their $\eta$-product as in Eq.\ \eqref{E4:eq6}
will be independent of time.  Furthermore, if another
operator $Z$ is independent of time and commutes with $H(t)$ for all
times, then also $(\Phi_1(t);Z\Phi_2(t))$ is a constant
in time.

     Suppose now that we have obtained a complete set of
solutions of Eq.\ \eqref{E4:eq18} across any desired time interval
$t_-\leq t\leq t_+$; equivalently, we have
for each closed interval $[t_-,t_+]$ in time
a linear operator $\Upsilon(t_+,t_-)$ such that
\begin{equation}\label{E4:eq19}
\Phi(t_+)\ =\ \Upsilon(t_+,t_-)\Phi(t_-),
\end{equation}
for any initial $\Phi(t_-)\in{\mathcal H}$.
One can show that $\Upsilon(t_+,t_-)$ is pseudounitary.
We define the input to the physical process taking place
to be the blocked vector
\begin{equation}\label{E4:eq20}
\Phi_{\rm in}(t_-,t_+)\ =\ \left[\begin{matrix}
(P^F\Phi(t_-))|_{{\mathcal H}^F}\\ (P^B\Phi(t_+))|_{{\mathcal H}^B}
\end{matrix}\right],
\end{equation}
and the output to be
\begin{equation}\label{E4:eq21}
\Phi_{\rm out}(t_+,t_-)\ =\ \left[\begin{matrix}
(P^F\Phi(t_+))|_{{\mathcal H}^F}\\ (P^B\Phi(t_-))|_{{\mathcal H}^B}
\end{matrix}\right].
\end{equation}
One can now show that, following Eq.\ \eqref{E4:eq17},
the operator $\tilde U(\Upsilon(t_+,t_-))$ is unitary and that
\begin{equation}\label{E4:eq22}
\Phi_{\rm out}(t_+,t_-)\ =\ \tilde U(\Upsilon(t_+,t_-))
\Phi_{\rm in}(t_-,t_+).
\end{equation}
In analyzing any physical process taking place in the interval 
$[t_-,t_+]$ we assume that the input state is given, known, or
controllable.  We can obviously multiply $\Phi(t)$ for all $t$
by a constant factor, with the desired outcome that
\begin{equation}\label{E4:eq23}
\langle\Phi_{\rm in}(t_-,t_+),\Phi_{\rm in}(t_-,t_+)\rangle\ =\ +1,
\end{equation}
(note the use of the Hilbert space norm), so that 
Eq.\ \eqref{E4:eq22} implies
\begin{equation}\label{E4:eq24}
\langle\Phi_{\rm out}(t_+,t_-),\Phi_{\rm out}(t_+,t_-)\rangle\ =\ +1.
\end{equation}
We therefore have in our possession
the bare bones of a probability interpretation
for the proposed scheme of kinematics and dynamics.

   We remark that the above interpretation as to what constitutes the
input and what the output to a dynamical process requires modification
if one or both ends of the time interval diverge:  since the
Hamiltonian can have nonreal eigenvalues, care must be taken
to avoid the divergent solutions associated with these closed-channel states.
A class of such problems is dealt with in the $S$-matrix formalism
of the following section.

     We continue to use the assumptions of the previous paragraph,
including the normalization condition Eq.\ \eqref{E4:eq23}
on the input.  Let $Z(t)$ be a pseudo-Hermitean operator,
and define the expectation value $[Z(t)]_{\rm Av}$ of
of $Z(t)$ for each
$t$, with the system in the state
$\Phi(t)$, in the standard manner:
\begin{equation}\label{E4:eq25}
[Z(t)]_{\rm Av}\ =\ (\Phi(t);Z(t)\Phi(t)).
\end{equation}
If $Z(t)=I$, the expectation
value is just the conserved
$\eta$-norm of $\Phi(t)$, which can be anywhere
between $+1$ and $-1$.  As mentioned above, we 
shall refer to this quantity as the net probability
current at time=$t$ in space-time.
This current is more closely analogous to an electric charge
than to a spatial electric current:  The electric charge is
the integrated value of the zeroth (time) component of the
four-vector electric current density over a surface t=constant;
looked in this way, a total electric charge amounts to a net
electric current crossing a complete space-like surface.
(Looked at another way, the state vector stays put at any given
time;  it is we who are moving through time, and hence we see
a changing state vector and thereby net currents of physical
quantities as probability, electric charge, etc.)
In the present case we do not define a four-vector
probability current density, but simply take
as a physical axiom that what is normally called ``probability''
is now to be regarded as the net probability current associated with a
quantum state at a given time.
For a general pseudo-Hermitean $Z(t)$ its expectation
value with respect to $\Phi(t)$
will be real, and will be taken to have the physical
meaning of the net current, or flow, or transport,
of the physical quantity associated
with $Z(t)$ across the chosen complete space-like surface,
as that surface moves forward in time with velocity $+1$.
Note that since the metric or velocity operator $\eta$ can have only
the dimensionless eigenvalues $+1$ and $-1$, a (likely unphysical)
density that gives rise to the current associated with
a $Z(t)$, which might take the form of $\langle\Phi(t),Z(t)\Phi(t)\rangle$,
and the expectation value itself, have the same physical dimensions.

\section{Formal scattering theory} \label{S:sec3}

    In this section we shall develop a theory of scattering patterned
after the developments in Levine \cite{R:Levine1}, Ch.\ 2.5, and
Newton \cite{R:Newton1}, Secs.\ 16.2 and 16.3.  
This formalism treats the time and energy coordinates in a different way
than it treats space and momentum, such that it
applies whether or not the underlying dynamics
is relativistically invariant;  a corresponding disadvantage is
the resulting lack of manifest relativistic invariance
of the terms in the perturbation theory expansion for the $S$ matrix
in cases as the field theory of Sec.\ \ref{S:sec4}, which is
there shown to be relativistically invariant in its Hamiltonian form.
The formalism will generalize the conventional one
in two respects:  first, evolution in both directions
in time will be included, and second, the zero$^{\rm th}$-order
Hamiltonian will be permitted to have some nonreal
eigenvalues in its spectrum---these correspond to 
asymptotically closed channels.  

     Suppose that the Hamiltonian is time-independent and
has the form
\begin{equation}\label{E5:eq1}
H\ =\ H^{[0]}\,+\,H^{[1]},
\end{equation}
where for both $\sigma=0,1$
\begin{equation}
H^{[\sigma]}\ =\ \left[\begin{matrix}
H^{[\sigma]FF} & -H^{[\sigma]BF\dagger} \\ 
H^{[\sigma]BF} & H^{[\sigma]BB} \end{matrix}\right],
\label{E5:eq2}
\end{equation}
where the diagonal-block operators are Hermitean, and where
$H^{[\sigma]BF}$ is unrestricted within the
bounds of physical reasonableness.
We shall adopt the picture that for large negative times and large
positive times the effects of $H^{[1]}$ are negligible:
the very early, as well as the very late, quantum state can be envisioned
as comprising superpositions of states, each of which describes two
spatially widely separated wave packets, such that each packet represents
an entity that does not interact with its partner, and the overall state is
a solution of the Schr\"odinger equation with $H^{[0]}$ as the
Hamiltonian.  

     We shall argue first that for nontransient transitions
to take place, the real sector of the eigenvalue
spectrum of the unperturbed Hamiltonian must, in effect, be positive
for both the states of forward motion in time (FMT)
and the states of backward motion in time (BMT).  In the
Hamiltonian given above, both $H^{[0]}$ and $H^{[1]}$
are to be time-independent; hence energy is conserved---there
will be a delta-function in overall energy that arises in the
results below.  A time-independent Hamiltonian that gives rise to 
nontransient transitions
between FMT and BMT states therefore requires that these two
sets of states have the real sectors of their respective eigenvalue 
spectra overlap.   We therefore abandon the picture that
states in BMT correspond to negative energy states:  in general,
both FMT and BMT states with real energy eigenvalues will be assumed
to have positive energies, or at least energies 
that are bounded below but not above along the real axis.

     Next, we specialize $H^{[0]}$ and $\eta$ so that they
have the properties that, although
the state space may be infinite-dimensional, permit them to be
jointly reduced to that
canonical form for finite-dimensional
matrices---described in Gohberg, et al. 
\cite{R:Gohberg1}, p.\ 33, Theorem 3.3---which
occurs when the minimal polynomial for the Hamiltonian matrix is
a product of distinct linear factors.  In particular, we assume that the
eigenstates of $H^{[0]}$ form a complete, orthogonal set
with respect to the underlying Hilbert space. 
Explicitly, suppose that there is a direct-sum
decomposition of the full state space $\mathcal H$ such that
\begin{subequations}\label{E5:eq3}
\begin{align}
{\mathcal H}\ &=\ {\mathcal H}^{R}\oplus{\mathcal
H}^{N}\label{E5:eq3a}\\
{\mathcal H}^R\ &=\ {\mathcal H}^{R,F}\oplus{\mathcal
H}^{R,B}\label{E5:eq3b}\\
{\mathcal H}^N\ &=\ {\mathcal H}^{N,1}\oplus{\mathcal H}^{N,2}
\label{E5:eq3c}
\end{align}
\end{subequations}
and a basis compatible with this decomposition, such that
$H^{[0]}$ is diagonal, $\eta$ is diagonal in the
subspace ${\mathcal H}^{R}$
belonging to the real eigenvalue spectrum of $H^{[0]}$, and
$\eta$ is has a simple, block off-diagonal form
in the subspace ${\mathcal H}^N$
belonging to the nonreal eigenvalue spectrum of $H^{[0]}$.
Furthermore, all vectors in ${\mathcal H}^{R,F}$ 
(resp., ${\mathcal H}^{R,B}$)
are eigenstates of $\eta$ with eigenvalue $+1$ (resp., $-1$).
Since each nonreal eigenvalue must have a complex conjugate partner,
we can take ${\mathcal H}^{N,1}$ and
${\mathcal H}^{N,2}$ to be copies
of one another, the former being associated with nonreal
eigenvalues with negative imaginary part, the latter
with nonreal eigenvalues with positive imaginary part.
In matrix form, therefore, 
we have assumed the existence of an invertible linear transformation
$T$ taking $H^{[0]}$ expressed in an arbitrary basis into a basis
so that we have
\begin{subequations}\label{E5:eq4}
\begin{equation}\label{E5:eq4a}
T^{-1}H^{[0]}T\ =\ 
\textit{diag}(\Delta^{R,F},\Delta^{R,B},\Delta^{N,1},\Delta^{N,2}),
\end{equation}
\begin{equation}\label{E5:eq4b}
T^{\dagger}\eta T\ =\ \left[\begin{matrix}
I^{R,F}&0&0&0\\ 0& -I^{R,B} &0&0\\ 0&0&0&U^{N,12}\\
0&0& U^{N,21}&0\end{matrix}\right].
\end{equation}\end{subequations}
In the above, $\Delta^{R,F}$ and $I^{R,F}$ are a real diagonal and
the unit matrix, respectively, acting on ${\mathcal H}^{R,F}$;
$\Delta^{R,B}$ and $I^{R,B}$ are a real diagonal and
the unit matrix, respectively, acting on ${\mathcal H}^{R,B}$;
$\Delta^{N,1}$ is a diagonal matrix with diagonal elements
having negative imaginary parts, acting on
${\mathcal H}^{N,1}$;  $U^{N,21}$ is the
unitary mapping of ${\mathcal H}^{N,1}$ 
onto ${\mathcal H}^{N,2}$ that takes an eigenstate
with eigenvalue $\Lambda$ (having, we have assumed, 
$\Im(\Lambda)<0$) with respect
to $\Delta^{N,1}$
into a partner eigenstate having an eigenvalue $\Lambda^\ast$
with respect to $\Delta^{N,2}$;
and $U^{N,12}$ is the inverse of $U^{N,21}$, in that
\begin{subequations}\label{E5:eq5}
\begin{align}
U^{N,21}\ &=\ (U^{N,12})^\dagger,\label{E5:eq5a} \\
U^{N,21}U^{N,12}\ &=\ I^{N,2},\label{E5:eq5b}\\
U^{N,12}U^{N,21}\ &=\ I^{N,1},\label{E5:eq5c}\\
U^{N,21}\Delta^{N,1}U^{N,12}\ &=\ 
(\Delta^{N,2})^\dagger,\label{E5:eq5d}
\end{align}
\end{subequations}
where $I^{N,1}$ and $I^{N,2}$ are the unit matrices in the spaces
$\mathcal{H}^{N,1}$ and $\mathcal{H}^{N,2}$, respectively.
The Eqs.\ \eqref{E5:eq5} are in accord with the pseudo-Hermitean property
for $H^{[0]}$.

     We remark that the above special form of $H^{[0]}$  excludes all
so-called ``ghost'' states associated with a real eigenvalue, and
retains only the simplest case of ghost states associated with
a nonreal eigenvalue---see Nagy \cite{R:Nagy1}, p.\ 14, for
definitions, and Ref.\ \cite{R:Gohberg1}, p.\ 33,
Theorem 3.3 for the joint canonical form 
of a general pseudo-Hermitean matrix 
and the metric matrix in the finite-dimensional case.

     We want to find that Green's function for $H^{[0]}$ such that
both open-channel $(R,F)$ states and 
closed-channel $(N,1)$ states evolve forward in time, while
open-channel $(R,B)$ and closed-channel $(N,2)$ 
states evolve backward in time.
If we put $\hbar=1$, $G^{[0]}(t-t')$ should satisfy
\begin{equation}\label{E5:eq6}
\biggl[i\frac{\partial}{\partial t}\,-\,H^{[0]}\biggr]
G^{[0]}(t-t')\ =\ \delta(t-t').
\end{equation}
The desired solution is
\begin{multline}\label{E5:eq7}
G^{[0]}(t-t')\ = \ 
\mathit{diag}\bigl(-i\theta(t-t')\exp[-i(t-t')\Delta^{R,F}],\\
i\theta(t'-t)\exp[-i(t-t')\Delta^{R,B}],\\
-i\theta(t-t')\exp[-i(t-t')\Delta^{N,1}],\\
i\theta(t'-t)\exp[-i(t-t')\Delta^{N,2}]\bigr),
\end{multline}
where $\theta$ is the unit step function.  The Fourier
transform of $G^{[0]}$ is
\begin{equation}\label{E5:eq8}
\tilde G^{[0]}(E)\ =\ \int_{-\infty}^{+\infty}
\exp(isE)G^{[0]}(s)\,ds
\end{equation}
\begin{multline}\tag{\ref{E5:eq8}a}
=\ \mathit{diag}\bigl([(E+i\epsilon)I^{R,F}-\Delta^{R,F}]^{-1},
[(E-i\epsilon)I^{R,B}-\Delta^{R,B}]^{-1},\\
[E I^{N,1}-\Delta^{N,1}]^{-1},
[E I^{N,2}-\Delta^{N,2}]^{-1}\bigr).
\end{multline}
Adding $+i\epsilon$ (respectively, 
$-i\epsilon$) to $E$ effects the usual small 
displacement of the poles of the integrand down (respectively, up)
from the real axis in the complex $E$-plane 
when recovering $G^{[0]}(t-t')$ from $\tilde G^{[0]}(E)$;
no displacement is needed for poles off the real axis.
If there is a nonzero gap between the entire nonreal
spectrum and the real axis, a very small raising or lowering
of the nonreal spectrum in the $E$-plane will not affect the
result in this subspace of $\mathcal H$; 
in such a case, we can give an abbreviated formula for
$\tilde G^{[0]}(E)$, that is,
\begin{equation}
\tilde G^{[0]}(E)\ =\ (EI+i\epsilon\eta-H^{[0]})^{-1},\label{E5:eq9}
\end{equation}
where $I$ is the unit operator in $\mathcal H$.

     We shall now specify a complete, orthogonal (in the Hilbert
space sense) set of eigenfunctions of $H^{[0]}$.
Let ${\mathcal S}^{R,F}$ (respectively, ${\mathcal S}^{R,B}$)
denote the subset 
of real eigenvalues of $H^{[0]}$ such that the
corresponding eigenstates are also eigenstates of $\eta$
with eigenvalue $+1$ (respectively, $-1$).  Let ${\mathcal S}^{N,1}$
denote the set of those eigenvalues of $H^{[0]}$ having
negative imaginary part, and ${\mathcal S}^{N,2}$ be the
set of complex conjugate points of those in ${\mathcal S}^{N,1}$.  
We shall assume that
${\mathcal S}^{N,1}$ is, or can be approximated by,
a discrete spectrum;
conceivably, however, there may exist $H^{[0]}$'s such that the
corresponding set ${\mathcal S}^{N,1}$ has a nondiscrete topology,
e.g., a subset of a curve in $\mathbb C$.

     We denote a state in the basis leading to the matrix
form of Eq.\ \eqref{E5:eq4} as $\Psi^{[0]Z,Y}_{\Lambda\gamma}$.
The index
$Z$ can take the values $R$ or $N$, and for $Z=R$, 
$Y$ can take the values $F$ or $B$, while for $Z=N$, $Y$ can take the values
$1$ or $2$.  Let
$\alpha_Y$ be defined as
\begin{equation}\label{E5:eq9.5}
\alpha_Y\ =\ 
\begin{cases}
+1  &\text{if $Y=F$,}\\
-1, &\text{if $Y=B$,}
\end{cases}
\end{equation}
$\Lambda$ be the eigenvalue
of $H^{[0]}$, and $\gamma$ (an
index which is implicitly dependent
on the other quantum numbers) label degenerate states
with respect to $H^{[0]}$.   We note the following
behavior of these eigenstates under the action of $\eta$:
\begin{subequations}\label{E5:eq10}
\begin{align}
\eta\Psi^{[0]R,Y}_{\Lambda,\gamma}\ &=\ \alpha_Y\Psi^{[0]R,Y}_{\Lambda,\gamma}
\label{E5:eq10a}\\
\eta\Psi^{[0]N,1}_{\Lambda\gamma}
\ &=\ \Psi^{[0]N,2}_{\Lambda^\ast\gamma},\label{E5:eq10b}\\
\eta\Psi^{[0]N,2}_{\Lambda^\ast\gamma}
\ &=\ \Psi^{[0]N,1}_{\Lambda\gamma}.\label{E5:eq10c}
\end{align}\end{subequations}
The Hilbert space orthonormality of the states and the completeness
relation are as follows:
\begin{subequations}\label{E5:eq11}
\begin{align}
&\bigl(\Psi^{[0]Z',Y'}_{\Lambda'\gamma'}\bigr)^\dagger 
\bigl(\Psi^{[0]Z,Y}_{\Lambda\gamma}\bigr)
\ =\ \delta^{Z'Z}\delta^{Y'Y}
\begin{cases} \delta(\Lambda'-\Lambda)\delta_{\gamma'\gamma}& 
\text{if $Z=R$,}\\ \delta_{\Lambda'\Lambda}\delta_{\gamma'\gamma}
&\text{if $Z=N$,}\end{cases}\label{E5:eq11a}\\
&I\ =\ \sum_{Y=F,B}\int_{E\in{\mathcal S}^{R,Y}}\sum_{\gamma}
\Psi^{[0]R,Y}_{E\gamma}\bigl(\Psi^{[0]R,Y}_{E\gamma}\bigr)^\dagger
dE \nonumber\\
&\qquad+\sum_{\Lambda\in{\mathcal S}^{N,1}}\sum_{\gamma}\left[
\Psi^{[0]N,1}_{\Lambda\gamma}
\bigl(\Psi^{[0]N,1}_{\Lambda\gamma}\bigr)^\dagger
\,+\,
\Psi^{[0]N,2}_{\Lambda^\ast\gamma}
\bigl(\Psi^{[0]N,2}_{\Lambda^\ast\gamma}\bigr)^\dagger
\right]
\label{E5:eq11b}
\end{align}
\end{subequations}

     We find the time-dependent, open-channel solutions of the
Schr\"odinger equation with $H^{[0]}$ as Hamiltonian to be
\begin{equation}\label{E5:eq12}
\Phi^{[0]R,Y}_{E\gamma}(t)\ =\ \exp(-iEt)\Psi^{[0]R,Y}_{E\gamma}.
\end{equation}
Then the full scattering wave function $\Phi^{R,Y}_{E\gamma}(t)$
with input as $\Phi^{[0]R,Y}_{E\gamma}(t)$ for $t\to -\alpha_Y\infty$
satisfies the integral equation
\begin{equation}
\Phi^{R,Y}_{E\gamma}(t)
\ =\ \Phi^{[0]R,Y}_{E\gamma}(t)
\,+\,\int_{-\infty}^{+\infty}G^{[0]}(t-t_1)H^{[1]}
\Phi^{R,Y}_{E\gamma}(t_1)dt_1.\label{E5:eq13}
\end{equation}
We presume that this equation can be solved by 
unlimited Neumann iterations, with the result
\begin{equation}\label{E5:eq14}
\begin{split}
\Phi^{R,Y}_{E\gamma}(t)
\ &=\ \Phi^{[0]R,Y}_{E\gamma}(t)
\,+\,\int_{-\infty}^{+\infty}G^{[0]}(t-t_1)H^{[1]}
\Phi^{[0]R,Y}_{E\gamma}(t_1)dt_1 \\
&\ +\sum_{j=2}^\infty\idotsint_{-\infty}^{+\infty}dt_1\cdots dt_j
G^{[0]}(t-t_1)H^{[1]}\\ &\qquad\times
\biggl[\prod_{k=2}^{j}G^{[0]}(t_{k-1}-t_{k})H^{[1]}
\biggr]\Phi^{[0]R,Y}_{E\gamma}(t_j).\end{split}
\end{equation}

     In the rhs of Eq.\ \eqref{E5:eq14}
let us now ({\it i}) use Eq.\ \eqref{E5:eq12} for the zero-order
wave functions, ({\it ii})
substitute the inverse of Eq.\ \eqref{E5:eq8}
for each entry $G^{[0]}(t_{k-1}-t_k)$ in the product in
Eq.\ \eqref{E5:eq14}, ({\it iii}) change variables of integration from
$t_k$ to $s_k$ (for $k=2,\ldots,j$, while $t_1$ is unchanged)
in the $j^{\text{th}}$ summand, where
\begin{subequations}
\begin{align}
&s_k\ =\ t_{k-1}-t_k, \ \ \text{for $k=2,\ldots,j$,}\label{E5:eq15a}\\
&\text{so that}\nonumber\\
&-t_j\ =\ -t_1+\sum_{k=2}^j s_k,\label{E5:eq15b}
\end{align}
\end{subequations}
({\it iv}) carry out the integrals over $s_2,\ldots,s_j$
in the $j^{\text{th}}$ summand, and ({\it v}) do the 
resulting integrals
involving delta-functions in energy.
We define the transition operator $T(E)$ as
\begin{subequations}\label{E5:eq16}
\begin{align}
T(E)\ &=\ H^{[1]}\sum_{j=0}^\infty\Bigl[\tilde G^{[0]}(E)H^{[1]}
\Bigr]^j\label{E5:eq16a} \\
&=\ H^{[1]}\Bigl[I -\tilde G^{[0]}(E)H^{[1]}\Bigr]^{-1}\label{E5:eq16b}\\
&=\ \Bigl[I -H^{[1]}\tilde G^{[0]}(E)\Bigr]^{-1}H^{[1]},\label{E5:eq16c}
\end{align}
\end{subequations}
where the zero power of an operator is the unit operator.
Then Eq.\ \eqref{E5:eq14} reduces to
\begin{equation}\label{E5:eq17}
\Phi^{R,Y}_{E\gamma}(t)
\ =\ \exp(-iEt)\Psi^{[0]R,Y}_{E\gamma}
\ +\ \int_{-\infty}^{+\infty} G^{[0]}(t-t_1)\exp(-iEt_1)T(E)
\Psi^{[0]R,Y}_{E\gamma}dt_1.
\end{equation}

     Let us now take the $\eta$-product of both sides of
Eq.\ \eqref{E5:eq17} with $\Phi^{[0]R,Y'}_{E'\gamma'}(t)$, while also
inserting the unit operator, in the form of the rhs of Eq.\
\eqref{E5:eq11b}, following the Green's function in the integrand
of Eq.\ \eqref{E5:eq17}:  using the expression Eq.\ \eqref{E5:eq7}
for the Green's function, we find, after some manipulation, that
\begin{equation}\label{E5:eq18}
\begin{split}
\bigl(\Phi&^{[0]R,Y'}_{E'\gamma'}(t);\Phi^{R,Y}_{E\gamma}(t)\bigr)
\ =\ \alpha_{Y'}\delta^{Y'Y}\delta(E'-E)\delta_{\gamma'\gamma}
\,-i\bigl(\Psi^{[0]R,Y'}_{E'\gamma'};\eta T(E)\Psi^{[0]R,Y}_{E\gamma}
\bigr)\\
&\times\Bigl[\delta^{Y'F}\int_{-\infty}^t
\exp[i(E'-E)t_1]dt_1\,+\,\delta^{Y'B}
\int_t^{+\infty}\exp[i(E'-E)t_1]dt_1\Bigr].
\end{split}
\end{equation}
The derivation from Eq.\ \eqref{E5:eq18}
of an expression for the transition probability per unit time is carried out
in an Appendix.
If we define the inverse function to Eq.\ \eqref{E5:eq9.5} as
\begin{equation}
\bar Y_\alpha\ =\ \begin{cases} F &\text{if $\alpha=+1$,}\\
B, &\text{if $\alpha=-1$,}\end{cases}\label{E5:eq19}
\end{equation}
then as $|t|\to\infty$, Eq.\ \eqref{E5:eq18} has the limiting forms
\begin{equation}\label{E5:eq20}
\begin{split}
\bigl(\Phi^{[0]R,Y'}_{E'\gamma'}(t)&;
\Phi^{R,Y}_{E\gamma}(t)\bigr)
\ \rightarrow\ \alpha_{Y'}\delta^{Y'Y}\delta(E'-E)\delta_{\gamma'\gamma}
-2\pi i\delta(E'-E)\\
&\times\delta^{Y'\bar Y_\alpha}
\bigl(\Psi^{[0]R,Y'}_{E'\gamma'};\eta T(E)\Psi^{[0]R,Y }_{E\gamma}
\bigr),
\text{\ as $t\to\alpha\infty$}.
\end{split}
\end{equation}

     We analyze Eq.\ \eqref{E5:eq20} to determine the analogs
of reflection and transmission coefficients, and assemble the
results into an $S$-operator.  That is, we want to have
\begin{equation}\label{E5:eq22}
\bigl(\Phi^{[0]R,Y '}_{E'\gamma'}(t);\Phi^{R,Y }_{E\gamma}(t)\bigr)
\to\begin{cases}\bigl(\Psi^{[0]R,Y '}_{E'\gamma'};S
\Psi^{[0]R,Y }_{E\gamma}\bigr),&
\text{for $t\to+\alpha_{Y'}\infty$,}\\
\bigl(\Psi^{[0]R,Y'}_{E'\gamma'};
\Psi^{[0]R,Y}_{E\gamma}\bigr),&
\text{for $t\to-\alpha_{Y'}\infty$.}\end{cases}
\end{equation}
On the basis of a comparision of Eqs.\ \eqref{E5:eq20} and
\eqref{E5:eq22},  we proceed to define the $S$-operator as
an entity that acts on, and only on, the subspace 
${\mathcal H}^R$ of $\mathcal H$.  We define $I^R$
as the identity operator within ${\mathcal H}^R$, and
$X^{RR}$ as the restriction of a general operator 
$X:{\mathcal H}\to{\mathcal H}$ to the suboperator that maps
to ${\mathcal H}^R\to{\mathcal H}^R$.  We note that
in the special cases treated here of an $H^{[0]}$ and $\eta$ of the form of
Eq.\ \eqref{E5:eq4},
$H^{[0]RR}$ and $\eta^{RR}$ commute, and correspondingly
$H^{[0]RR}$ is Hermitean.  Then if we let
\begin{equation}
S\ =\ I^R\,-\,2\pi i\int_{-\infty}^{+\infty}dE
[\delta(EI-H^{[0]})\eta T(E)\delta(EI-H^{[0]})]^{RR},
\label{E5:eq23}
\end{equation}
Eqs.\ \eqref{E5:eq20} and \eqref{E5:eq22} are in accord.

     It remains to prove that the $S$-operator
acts unitarily within ${\mathcal H}^R$.  In fact, we infer from
Eq.\ \eqref{E5:eq23} that
\begin{equation}
SS^\dagger\ -\ I^R\ =\ -2\pi i\int_{-\infty}^{+\infty}dE
[\delta(EI-H^{[0]})\eta]^{RR}[\Xi(E)]^{RR}[\eta\delta(EI-H^{[0]}]^{RR},
 \label{E5:eq24}
\end{equation}
where, by definition, 
\begin{equation}
\Xi(E)\ =\ T(E)\eta - \eta T(E)^\dagger+2\pi i
T(E)\delta(EI-H^{[0]})T(E)^\dagger;\label{E5:eq25}
\end{equation}
we made use of the properties that the operator $\delta(EI-H^{[0]})$ has
its cokernel and image contained within the subspace ${\mathcal H}^R$,
and that, as a result of Eqs.\ \eqref{E5:eq4a} and \eqref{E5:eq4b},
$\delta(E-H^{[0]})$ is Hermitean as well as pseudo-Hermitean.
We want to prove that $\Xi(E)$ equals the zero operator
for all real values of $E$.  To do this, we modify the argument that
leads to Ref.\ \cite{R:Levine1}, Eq.\ (5.29).  The steps are
very similar, except that now $\eta H^{[1]\dagger}=H^{[1]}\eta$, and
we need the easily verified result
\begin{equation}
\tilde G^{[0]}(E)\eta -\eta\tilde G^{[0]}(E)^\dagger\ =\ -2\pi i
\delta(\eta E-\eta H^{[0]})\ =\ -2\pi i\delta(E-H^{[0]}).\label{E5:eq26}
\end{equation}
A similar proof can be constructed to show that $S^\dagger S=I^R$.

    We remark finally that, in a rigorous
analysis of a physical process in a
finite time interval (e.g., in a quantum measurement theory),
it will be necessary to include the
closed-channel states due to the incompleteness of the open-channel
states in the Hilbert space.

\section{Direct sum of two quantised real scalar fields} \label{S:sec4}

     In this section we shall advance a dynamics
for a quantum field theory, the state space
of which comprises the direct sum of the state spaces for two
quantized real (i.e., Hermitean) Klein-Gordon fields.  We shall
show that the dynamics is relativistically invariant, and work out
a simple example of collision dynamics using 
first-order perturbation theory.

    We use the time and space coodinates $x^\mu=(x^0,x^1,x^2,x^3)$,
the metric tensor $g_{\mu\nu}=\text{diag}(+1,-1,-1,-1)$,
the Hilbert space notation of Sec.\ \ref{S:sec2}, and take both
$\mathcal{H}^F$ and $\mathcal{H}^B$ to be copies of
Fock space (see Schweber \cite{R:Schweber2}, Ch.\ 7a) for an
electrically neutral spin zero particle of mass $m$.
As before, $Y$ can take either value $F$ or $B$.
Let the zero state be
$\Upsilon(Y,z)\in\mathcal{H}^Y$, the vacuum state (with Hilbert space
norm $+1$) be 
$\Upsilon(Y,0)\in\mathcal{H}^Y$, and let $a_{\mathbf{p}}^Y$
and $a_{\mathbf{p}}^{Y\dagger}$ be the operator 
that destroys
and, respectively, creates a particle of 3-momentum $\mathbf{p}$.
We normalize these operators such that their commutators are
\begin{subequations}\label{E6:eq1}
\begin{align}
[a_{\mathbf{p}}^Y,a_{\mathbf{p}'}^Y]\ &\ =\ 0,\label{E6:eq1a}\\
[a_{\mathbf{p}}^{Y\dagger},
a_{\mathbf{p}'}^{Y\dagger}]\ &\ =\ 0,\label{E6:eq1b}\\
[a_{\mathbf{p}}^{Y},
a_{\mathbf{p}'}^{Y\dagger}]\ &\ =\ \delta^3(\mathbf{p}-\mathbf{p}')
I^Y.\label{E6:eq1c}
\end{align}\end{subequations}
Now let $\{\mathbf{p}_1,\mathbf{p}_2,\ldots,\mathbf{p}_N\}$
be a finite set of distinct 3-momenta;  then
we have the following state in $\mathcal{H}^Y$
that corresponds to one particle with 3-momentum
$\mathbf{p}_1$,\ldots, and
one particle with 3-momentum $\mathbf{p}_N$:
\begin{equation}\label{E6:eq2}
\Upsilon(Y,N;\mathbf{p}_1,\mathbf{p}_2,\ldots,\mathbf{p}_N)
\ =\ (N!)^{-1/2}a_{\mathbf{p}_1}^{Y\dagger}\cdots
a_{\mathbf{p}_N}^{Y\dagger}\Upsilon(Y,0).
\end{equation}
The normalization guarantees that $I_N^Y$,
\begin{equation}\label{E6:eq2X}
I_N^Y\ =\ \idotsint_{\mathbb{R}^3}d^3p_1\cdots d^3p_N
\Upsilon(Y,N;\mathbf{p}_1,\mathbf{p}_2,\ldots,\mathbf{p}_N)
\Upsilon(Y,N;\mathbf{p}_1,\mathbf{p}_2,\ldots,\mathbf{p}_N)^\dagger
\end{equation}
is a projection operator from
$\mathcal{H}^Y$ to the subspace of $N$-particle states in $\mathcal{H}^Y$.

     Let $U^{FB}$ be the simple linear
mapping from $\mathcal{H}^B$ to $\mathcal{H}^F$, in that
\begin{subequations}\label{E6:eq3}
\begin{align}
U^{FB}\Upsilon(B,z)\ &=\ \Upsilon(F,z),\\
U^{FB}\Upsilon(B,0)\ &=\ \Upsilon(F,0),\\
U^{FB}\Upsilon(B,1;\mathbf{p}_1)
\ &=\ \Upsilon(F,1;\mathbf{p}_1),\\
&\vdots\ldots\ .\nonumber
\end{align}
\end{subequations}
The operator $U^{FB}$ 
obviously has a two-sided inverse $U^{BF}$
that coincides with its adjoint, i.e.,
\begin{subequations}\label{E6:eq4}
\begin{align}
U^{BF}\ =\ (U^{FB}&)^{-1}\ =\ (U^{FB})^\dagger,\\
U^{FB} U^{BF}\ &=\ I^F,\\
U^{BF} U^{FB}\ &=\ I^B.
\end{align}
\end{subequations}

     We reconstruct the subfields in terms of the destruction
and creation operators in the manner of
Peskin and Schroeder \cite{R:Peskin1}, p.\ 21, with
$\omega_{\bf{p}}=\sqrt{k^2+m^2}>0$:
\begin{subequations}\label{E6:eq5}
\begin{align}
\phi^Y(\mathbf{x})\ &=\ \int_{\mathbb{R}^3}
\frac{d^3p}{[2\omega_{\mathbf{p}}(2\pi)^3]^{1/2}}
\Bigl[a^Y_{\mathbf{p}}\exp(i\mathbf{p}\cdot\mathbf{x})
\,+\,a^{Y\dagger}_{\mathbf{p}}\exp(-i\mathbf{p}\cdot\mathbf{x})
\Bigr],\label{E6:eq5a}\\
\pi^Y(\mathbf{x})\ &=\ \int_{\mathbb{R}^3}d^3p
\frac{(-i)(\omega_{\mathbf{p}})^{1/2}}{[2(2\pi)^3]^{1/2}}
\Bigl[a^Y_{\mathbf{p}}\exp(i\mathbf{p}\cdot\mathbf{x})
\,-\,a^{Y\dagger}_{\mathbf{p}}\exp(-i\mathbf{p}\cdot\mathbf{x})
\Bigr].\label{E6:eq5b}
\end{align}
\end{subequations}
The latter entities have the commutators
\begin{subequations}\label{E6:eq6}
\begin{align}
[\phi^Y(\mathbf{x}),\phi^Y(\mathbf{y})]\ &=\ 0,\label{eq6a}\\
[\pi^Y(\mathbf{x}),\pi^Y(\mathbf{y})]\ &=\ 0,\label{eq6b}\\
[\phi^Y(\mathbf{x}),\pi^Y(\mathbf{y})]\ &=\ i
\delta^3(\mathbf{x}-\mathbf{y})I^Y.\label{eq6c}
\end{align}\end{subequations}
The field operators satisfy
\begin{subequations}\label{E6:eq7}
\begin{align}
\phi^B(\mathbf{x})
\ &=\ U^{BF}\phi^F(\mathbf{x})U^{FB},\\
\pi^B(\mathbf{x})
\ &=\ U^{BF}\pi^F(\mathbf{x})U^{FB}.
\end{align}
\end{subequations}
The physical dimensions of the 
fields $\phi^Y(\mathbf{x})$ and $\pi^Y(\mathbf{x})$
are (length)$^{-1}$ and (length)$^{-2}$, respectively,
modulo powers of $\hbar$ and $c$.

     We shall now formulate a particular case of dynamics and 
show how to verify
that the theory is relativistically invariant.  Following the pattern 
in Ref.\ \cite{R:Peskin1}, Eqs.\ (2.8), (2.18), (2.19), and (4.12), 
we postulate
{\it ad hoc} the following operator for the energy density
$T^{00}(\mathbf{x})$:
\begin{equation}\label{E6:eq8}
T^{00}(\mathbf{x})\ =\ T^{[0]00}(\mathbf{x})\,+\,T^{[1]00}(\mathbf{x}),
\end{equation}
where
\begin{subequations}\label{E6:eq9}
\begin{align}
T^{[0]00}(\mathbf{x})\ &=\ \left[
\begin{matrix}\frac{1}{2}\bigl[\pi^F(\mathbf{x})^2
+\mathbf{\nabla_x}\phi^F\cdot\mathbf{\nabla_x}\phi^F
+m^2\phi^F(\mathbf{x})^2\bigr] & 
\\ 0 & \end{matrix}\right.\nonumber\\
&\qquad\qquad\left.\begin{matrix} & 0\\ &
\frac{1}{2}\bigl[\pi^B(\mathbf{x})^2
+\mathbf{\nabla_x}\phi^B\cdot\mathbf{\nabla_x}\phi^B
+m^2\phi^B(\mathbf{x})^2\bigr]
\end{matrix}\right],\label{E6:eq9a}
\\ & \nonumber \\
T^{[1]00}(\mathbf{x})\ &=\ \left[\begin{matrix}
\frac{1}{4}\zeta^F\phi^F(\mathbf{x})^4 &
 -\frac{1}{4}\xi\phi^F(\mathbf{x})^2 
U^{FB}\phi^B(\mathbf{x})^2 \\
 \frac{1}{4}\xi\phi^B(\mathbf{x})^2 
U^{BF}\phi^F(\mathbf{x})^2 &
\frac{1}{4}\zeta^B\phi^B(\mathbf{x})^4 
\end{matrix}\right].\label{E6:eq9b}
\end{align}\end{subequations}
The dimensionless coupling constants 
$\zeta^F\geq 0$, $\zeta^B\geq 0$, and (following, 
if needed, a separate phase transformation
in ${\mathcal H}^F$ and ${\mathcal H}^B$) $\xi$,
are all real.  The Hamiltonian is defined as follows:
\begin{subequations}\label{E6:eq10}
\begin{align}
H\ &=\ H^{[0]}\,+\,H^{[1]},\\
H^{[0]}\ &=\ \int_{\mathbb{R}^3} T^{[0]00}(\mathbf{x})\,d^3x,
\label{E6:eq10a}\\
H^{[1]}\ &=\ \int_{\mathbb{R}^3} T^{[1]00}(\mathbf{x})\,d^3x.
\label{E6:eq10b}
\end{align}
\end{subequations}
The momentum-density operator $T^{j0}(\mathbf{x})$ and momentum operator
$\Pi^j$ have a form that does not involve the interaction
coupling parameters $\zeta^F$, $\zeta^B$, or $\xi$:
\begin{equation}\label{E6:eq11}
T^{j0}(\mathbf{x})\ =\ \left[\begin{matrix}
-\frac{1}{2}\bigl[\pi^F(\mathbf{x})
\frac{\partial\phi^F}{\partial x^j}+
\frac{\partial\phi^F}{\partial x^j}\pi^F(\mathbf{x})\bigr]
& 0 \\ 0 &
-\frac{1}{2}\bigl[\pi^B(\mathbf{x})
\frac{\partial\phi^B}{\partial x^j}+
\frac{\partial\phi^B}{\partial x^j}\pi^B(\mathbf{x})\bigr]
\end{matrix}\right],
\end{equation}
\begin{equation}\label{E6:eq12}
{\Pi}^j\ =\ \int_{\mathbb{R}^3}T^{j0}(\mathbf{x})\,d^3x.
\end{equation}
The energy-flow operator $T^{0j}(\mathbf{x})$ 
is taken to be the same operator as $T^{j0}(\mathbf{x})$.
The stress-tensor operator $T^{jk}(\mathbf{x})$ is chosen
as follows:
\begin{equation}\label{E6:eq12x1}
T^{jk}(\mathbf{x})\ =\ T^{[0]jk}(\mathbf{x})\,+\,T^{[1]jk}(\mathbf{x}),
\end{equation}
where
\begin{subequations}\label{E6:eq12x2}
\begin{align}
T^{[0]jk}(\mathbf{x})\ &=\ \left[
\begin{matrix}
\frac{\partial\phi^F}{\partial x^j}
\frac{\partial\phi^F}{\partial x^k}
+(1/2)\delta^{jk}\bigl[\pi^F(\mathbf{x})^2
-\mathbf{\nabla_x}\phi^F\cdot\mathbf{\nabla_x}\phi^F
-m^2\phi^F(\mathbf{x})^2\bigr] & 
\\ 0 & \end{matrix}\right.\nonumber\\
&\qquad\left.\begin{matrix} & 0\\ &
\frac{\partial\phi^B}{\partial x^j}
\frac{\partial\phi^B}{\partial x^k}
+(1/2)\delta^{jk}\bigl[\pi^B(\mathbf{x})^2
-\mathbf{\nabla_x}\phi^B\cdot\mathbf{\nabla_x}\phi^B
-m^2\phi^B(\mathbf{x})^2\bigr] 
\end{matrix}\right],\label{E6:eq12x3}
\\ & \nonumber \\
T^{[1]jk}(\mathbf{x})\ &=\ \left[\begin{matrix}
-\frac{1}{4}\delta^{jk}\zeta^F\phi^F(\mathbf{x})^4 &
 \frac{1}{4}\xi\delta^{jk}\phi^F(\mathbf{x})^2 
U^{FB}\phi^B(\mathbf{x})^2 \\
 -\frac{1}{4}\delta^{jk}\xi\phi^B(\mathbf{x})^2 
U^{BF}\phi^F(\mathbf{x})^2 &
-\frac{1}{4}\delta^{jk}\zeta^B\phi^B(\mathbf{x})^4 
\end{matrix}\right].\label{E6:eq12x4}
\end{align}\end{subequations}

     We can now define several other operators on the space of
time-dependent states, to assemble a set
of generators for the Poincar{\'e} group and Schr{\"o}dinger equation:
\begin{subequations}\label{E6:eq13}
\begin{align}
{\Pi}^0\ &=\ \frac{1}{i}\frac{\partial}{\partial x^0}I,\\
\Omega\ &=\ {\Pi}^0\,+\,H,\\
L^j\ &=\ \int_{\mathbb{R}^3}
\epsilon^{jkl}x^k T^{l0}(\mathbf{x})\,d^3x,\\
B^j\ &=\ x^0\Pi^j\,-\,\int_{\mathbb{R}^3}x^jT^{00}(\mathbf{x})d^3x
\end{align}
\end{subequations}
The rotation generators $L^j$ and Lorentz ``boost'' generators 
$B^j$ have been defined as in
Ref.\ \cite{R:BjD2}, Eqs.\ (11.57) and (15.19).
We call $\Omega$ the Schr{\"o}dinger operator, as the 
Schr{\"o}dinger equation for the time-dependent state 
$\Phi(x^0)\in\mathcal{H}^F\oplus\mathcal{H}^B$ is
\begin{equation}\label{E6:eq14}
\Omega \Phi(x^0)\ =\ 0.
\end{equation}
The real linear span of the set $\mathcal{P}$ of ten operators
\begin{equation}\label{E6:eq15}
\mathcal{P}\ =\ \{ -H,\, \{ {\Pi}^j, L^j, B^j,
\ \text{for}\ j=1,2,3\}\}
\end{equation}
comprises a Lie algebra that is
isomorphic to that of the Poincar{\'e} group, as
is verified by computing the following commutators (we
omit the calculational details):
\begin{subequations}\label{E6:eq16}
\begin{align}
[\Pi^j,H]\ &=\ 0,\\ 
[L^j,H]\ &=\ 0,\\
[B^j,H]\ &=\ -i\Pi^j,\\
[\Pi^j,\Pi^k]\ &=\ 0,\\
[L^j,\Pi^k]\ &=\ i\epsilon^{jkl}\Pi^l,\\
[B^j,\Pi^k]\ &=\ -i\delta^{jk}H,\\
[L^j,L^k]\ &=\ i\epsilon^{jkl}L^l,\\
[L^j,B^k]\ &=\ i\epsilon^{jkl}B^l,\\
[B^j,B^k]\ &=\ -i\epsilon^{jkl}L^l.
\end{align}\end{subequations}
We also have the following commutators with $\Pi^0$:
\begin{subequations}\label{E6:eq17}
\begin{align}
[\Pi^0,H]\ &=\ 0,\\
[\Pi^0,\Pi^j]\ &=\ 0,\\
[\Pi^0,L^j]\ &=\ 0,\\
[\Pi^0,B^j]\ &=\ -i\Pi^j.
\end{align}\end{subequations}

     It proves to be the case that $\Omega$ commutes with all ten
basis elements and hence all elements of the Poincar{\'e} group's
Lie algebra:
\begin{equation}\label{E6:eq18}
[X,\Omega]\ =\ 0,\ \textrm{for all}\ X\in\mathcal{P},
\end{equation} 
and all elements of the component of the identity of the Poincar\'{e}
group are obtained by exponentiating $(-i)$ times some element of the
Lie algebra.
Hence, the dynamics entailed by the Schr{\"o}dinger equation
is invariant under the 
component of the identity
of the Poincar{\'e} group, in the sense
that the application of any
element of the group to a solution of the equation of motion
yields a transformed state that
is also a solution to the same equation of motion, i.e., 
Eq.\ \eqref{E6:eq14}.  We shall not consider the discrete
transformations of space and time 
herein, except to note that, since in general
$\zeta^F\neq\zeta^B$,  time reversal---in the 
strict sense of a simple
interchange of FMT and BMT---need not be 
a symmetry of the above dynamics;
the latter assertion should be distinguished from symmetry under
conventional time reversal, however, which is more accurately
termed ``reversal of the direction of [spatial] motion''---see 
Wigner \cite{R:Wigner1}, p.\ 325.  In the present context
a distinction between reversal of time and reversal of motion
can be meaningful, and thereby determine an absolute
direction of time---see Zeh \cite{R:Zeh1}, p.\ 3, footnote 1.
 
     A similarity transformation by the operator 
$W$ of an operator $X$ is defined as $WXW^{-1}$.  
We define the pseudounitary operator $W$ as
\begin{equation}\label{E6:eq19}
W\ =\ \left[\begin{matrix}
I^F\cosh\theta & U^{FB}\sinh\theta\\
U^{BF}\sinh\theta & I^B\cosh\theta
\end{matrix}\right],
\end{equation}
where $\theta$ is a real constant.  A similarity
transformation by $W$ leaves the rhs of
Eq.\ \eqref{E6:eq9a} unchanged, and transforms
the rhs of Eq.\ \eqref{E6:eq9b}
into another operator of the same form with different coupling 
constants.  If the discriminant  
\begin{equation}\label{E6:eq20}
D\ =\ (\zeta^F -\zeta^B)^2/4\,-\,\xi^2
\end{equation}
is positive, and we choose 
\begin{equation}\label{E6:eq21}
\theta\ =\ -(1/2)\mathrm{arctanh}[2\xi/(\zeta^F-\zeta^B)],
\end{equation}
the resultant operator is block diagonal, i.e.,
there is no coupling between
FMT and BMT (as redefined).  Hence we need
$\xi\neq 0$ and $D$ nonpositive to guarantee a nontrivial dynamics.
A simplification also occurs if both $D$ is negative and
\begin{equation}\label{E6:eq22}
\theta\ =\ -(1/2)\mathrm{arctanh}[(\zeta^F-\zeta^B)/2\xi].
\end{equation}
The modified coupling constants then have equal diagonal coefficients.

    A further remark:   In the above kinematics there is a
family of vacuum states given by 
$\alpha\Upsilon(F,0)\oplus\beta\Upsilon(B,0)$,
with $\alpha$ and $\beta$ being complex constants 
(at least one of which is nonzero) modulo
equivalence by an overall nonzero complex multiplier.
Hence the geometry of the space of rays of vacuum states is
$\mathbb{C}P^1$, which is homeomorphic to 
the Riemann sphere, i.e., $S^2$ (Frankel \cite{R:Frankel1}, p. 22).
This fact will be used in Sec.\ \ref{S:sec5}.

    To complete this section we shall apply first-order
perturbation theory to the above formalism to estimate the
cross section for an input state of two particles, 
both in FMT or both in BMT, to scatter into
an output state of two particles, where the two-particle
output may be either jointly in FMT or jointly in BMT.  First-order
perturbation theory consists in substituting $H^{[1]}$ for
$T(E)$ in Eq.\ \eqref{App:eq3}.  
After dropping several divergent self-energy terms, 
we find the result
is that given in Ref.\ \cite{R:Peskin1}, p.\ 112.
We work in the CM frame so that
the input particles have momentum $+\hat{\mathbf{p}}_{\rm{in}}|\mathbf{p}|$ 
and $-\hat{\mathbf{p}}_{\rm{in}}|\mathbf{p}|$, 
the output particles have
momentum $+\hat{\mathbf{p}}_{\rm{out}}|\mathbf{p}|$ 
and $-\hat{\mathbf{p}}_{\rm{out}}|\mathbf{p}|$, 
and so that the total energy $E_{\mathrm{CM}}$
and relative speed $v_{\mathrm{rel}}$ 
(as defined in Ref.\ \cite{R:Weinberg1}, Eq.\ (3.4.18)) are
\begin{subequations}\label{E6:eq23}
\begin{align} 
E_{\mathrm{CM}}\ &=\ 2\omega_{\mathbf{p}},\label{E6:eq23a}\\
v_{\mathrm{rel}}\ &=\  2|\mathbf{p}|/\omega_{\mathbf{p}}.\label{E6:eq23b}
\end{align}\end{subequations}
Then the total cross sections in ordinary units are
\begin{subequations}\label{E6:eq24}
\begin{align}
(\sigma_{\mathrm{total}})_{\mathrm{FMT}\leftarrow\mathrm{FMT}}
\ &=\ \frac{9(\zeta^F\hbar c)^2}
{8\pi E_{\mathrm{CM}}^2},\label{E6:eq24a}\\
(\sigma_{\mathrm{total}})_{\mathrm{BMT}\leftarrow\mathrm{BMT}}
\ &=\ \frac{9(\zeta^B\hbar c)^2}
{8\pi E_{\mathrm{CM}}^2},\label{E6:eq24b}\\
(\sigma_{\mathrm{total}})_{\mathrm{BMT}\leftarrow\mathrm{FMT}}
\ &=\ \frac{9(\xi\hbar c)^2}
{8\pi E_{\mathrm{CM}}^2}
\ =\ (\sigma_{\mathrm{total}})_{\mathrm{FMT}\leftarrow\mathrm{BMT}}.
\label{E6:eq24c}
\end{align}\end{subequations}

     By way of a numerical estimate, suppose that $\hbar c/E_{CM}$ is
half the pi meson Compton wavelength, that is about
$10^{-15}\,\rm{m}$, and that $\xi$
is about $10^{-10}$; $\zeta^F$ and $\zeta^B$ can be large, so
long as  $D\leq 0$ is satisfied.  The cross sections of
Eq.\ \eqref{E6:eq24c} are then about $10^{-50}\,\rm{m}^2$.  
These processes are sufficiently unlikely that
they are practically unobservable on a microscopic scale,
similar to most gravitation-induced phenomena.

     Note that a collision in which either
FMT$\leftarrow$FMT or BMT$\leftarrow$FMT can take place
will entail, on average, an apparent violation of conservation laws.
At a time earlier
than the collision, the quantum state appears to be
a superposition, or a kind of mixture, 
of FMT states and BMT states with 
equal total energies and momenta.  The small BMT component of the
state is part of the output, so that we do not, and by our rules
cannot, control this part of the temporally initial state.
This BMT component of the temporally earlier quantum state 
looks to our imagination
like a probability-amplitude wave converging---as our time 
increases---on the
collision event in space-time.  This wave interacts
very weakly with the constituents of
the local environment (the laboratory, the earth,
etc., all of which are in FMT),
even if this wave describes particles as $\pi^0$ mesons 
that, were they to appear
in an FMT state, would interact strongly with the same environment.
Hence to a first approximation we need not question
the fate of this output BMT wave in the past;
it will be effectively indetectible to us.  
(A collision and a detection amount to a second-order process.)
Nevertheless, the BMT $\pi^0$'s would presumably 
each decay into two BMT gamma rays at a time earlier
than the collision, which entities are not treated in the present
theory, but which would also interact weakly
with the FMT environment.  What would be observable
after the collision in an FMT laboratory is that 
there is a small probability that the input particles, including
all their energy and momentum, disappear.  There would thus be
an apparent nonconservation of energy and momentum,
as our instruments can conveniently detect only the
FMT part of the energy/momentum flow in space-time.
The observed stability of matter could be due to either
(1) the smallness of the FMT/BMT coupling, or (2) the circumstance
that in a hypothetical theory that describes the physical
world, fermion (lepton, baryon) quantum numbers 
associated with FMT and with BMT are separately conserved.

     To an extent, then, this theory gives a realization to the
popular picture of a time machine for travel
into the past, albeit only on the level of elementary
particle physics.  The process that a macroscopic entity
scatters coherently from an FMT state into a BMT state would be
improbable in the extreme.

\section{Further discussion and an application} \label{S:sec5}

     The physical picture that we have adopted amounts to saying that
the world can be described by a kinematics that looks like the
direct sum of the kinematics of two conventional quantum field
theories.  We propose the following visualization:  
The universe consists of a connected 
space-time manifold, within which
the ingredients of matter can be, besides in the conventional
range of FMT physical states, in BMT states;   
the dynamical coupling, that is the rate of quantum jumping,
of matter between these two sets of states is small, but nonzero.
What is of physical interest in the context of theory is establishing
criteria for determining if transitions
between the hypothetical set of BMT states and states in the known
FMT world occur at some very low level.  Aside from the
computation of scattering cross sections in Sec.\ \ref{S:sec4} and
remarks on vacuum states later in this section, we shall not deal with
this problem herein.  

     The formalism proposed in Sec.\ \ref{S:sec4}
presumes that particles in forward or backward motion have the same 
bare mass $m$;  the theory 
satisfies the criteria of relativistic invariance.
If we instead introduce distinct bare masses $m_F$ and $m_B$ in 
Eq.\ \eqref{E6:eq9a}, relativistic invariance fails.
A na\"{\i}ve consideration of the
possible theoretical structures does
not seem to exclude the possibility that the spectrum of
masses, spins, electric charges, etc., of elementary
particles could be widely different in the FMT and BMT
subspaces.  But
in another circumstance, Weinberg \cite{R:Weinberg1}, p.\ 145 made
the observation that the commutativity restriction for
the energy density operator at space-time points
separated by a nonzero space-like interval
is the ``$\ldots${\it condition that makes the combination of
Lorentz invariance and quantum mechanics so restrictive}''
(italics in the original).
There is not yet a counterpart to this condition in the
theory described here, as we have avoided
the introduction of a Heisenberg picture for field (or any) operators,
due in part to the fact that Hamiltonians can have
complex eigenvalues, and in part to nonlocal definitions
of input and output. 
The point we want to make is that relativistic invariance
may place severe restrictions on the possible mappings
from the state space and dynamics of one quantum field theory to that
of another, and thereby constrain the differences
between possible field
physics, and spectrum of
particle masses, spins, charges, etc.,
associated with the respective FMT and BMT sub-worlds.
This problem remains to be investigated.

          A proposal concerning the existence of matter that has an
internally reversed time sense was made by Stannard
\cite{R:Stannard1}.  The argument there was made in the 
context of the then-recently-discovered CP-noninvariance of
$K^0$-meson decays, and distinguishes the proposed new kind of 
matter (called ``Faustian'')
from conventional antimatter, which was described as ordinary matter
moving backwards in time.  There is a resemblance between the physics
of Stannard's Faustian matter and that of matter in BMT proposed
herein.   However, the article did not contain a mathematical
formulation of the equations of motion of such a generalized
system.  It may be said that the theory proposed herein is
a possible formulation of Stannard's hypotheses, accompanied
by the specifications ({\it i}) that the state space is
the direct sum, rather than a direct product,
of the state spaces of matter in FMT and in BMT, and
({\it ii}) that the quantum
state of the complete system is characterized by joint
forward and backward evolution or motion in time from a suitable
input combination of initial and final conditions.

     Feynman \cite{R:Feynman1} made an attempt 
to introduce negative probabilities into physics that
is distinct from the work 
cited in Nagy \cite{R:Nagy1} on indefinite metrics.  We emphasize
that in the theory presented here, probabilities are nonnegative
and S-matrices are unitary as opposed to pseudounitary.
The metric of indefinite sign is interpreted as giving rise to
a net current of something across a complete space-like surface, where
the current is associated with the probability in a
way that involves both input and output states.
Analogous to spin,  
the quantity that gives rise to the current is not further described, and
these flows are nonclassical:  the ``velocity'' of flow in space-time
can in effect have only the values $+1$ and $-1$, that is, the
eigenvalues of the metric operator $\eta$, corresponding
to FMT and BMT, respectively.  ``Current'',
``flow'', and ``transport'' in time
are taken as physically suggestive words, but we do not, and assert
that we need not, specify in the sense of classical mechanics 
either what it is that is flowing or the existence of any
extra parameter with respect to which the rate of flow
is defined.  The association of the expectation value
of a quantity with the net transport of that quantity
is taken as a physical axiom, which has no deeper
explanation in the present context. 

     We have introduced a theoretical construct in which
an event, taken as a cause, can have effects either earlier
or later than the cause, or both.  Concordantly, we 
adopt what is called
the ``block universe'' viewpoint by Price (\cite{R:Price1}, p.\ 12,
et seq.) and by Nahin (\cite{R:Nahin1}, p. 150, et seq.),
of the dynamically prescribed configuration of a system taken
as a whole for all space and for all times in a chosen interval.
An entity that can control the complete input to, and
observe the output from, such a system must in some sense
stand outside time and space as we know them, that is, 
must have what is called an ``atemporal Archimedean standpoint''
by Price (\cite{R:Price1}, p.\ 114).  This ``outside'' standpoint
is analogous to that in which an ordinary observer in space-time can
manipulate the input for solutions of the
steady-state, time-independent Schr\"{o}dinger equation.  
The phenomenon of closed causal chains,
in the sense of Reichenbach (\cite{R:Reichenbach1}, p.\ 36) or
Nahin (\cite{R:Nahin1}, p.\ 196) could arise in this
hypothetical universe.  Self-consistency of this process apparently
requires a kind of determinism, or a limitation on free will, that
is in contradiction to our present understanding.
The latter problem also arises in the hypothetical case of
topologically connected space-times with closed time-like
world lines---see Novikov (\cite{R:Novikov1}, p.\ 254) or
Nahin (\cite{R:Nahin1}, pp.\ 80--83).

     A conventional quantum field theory has a unique vacuum state,
a circumstance that permits simplifications, e.g., positioning
the energy axis so that the vacuum energy is zero.
In the field theory of Sec.\ \ref{S:sec4}, there are two vacuum states.
(We remark that the physical vacuum is also nonunique in 
some gauge theories---see, 
e.g. Kaku \cite{R:Kaku1}, Ch.\ 10---but this results from
assuming basic tachyon, or imaginary mass, 
fields with certain higher-than-second-order potential energy
terms in the classical field Lagrangian, such that the unique
mathematical vacuum is an unstable local maximum in the
field potential energy, and the minimum energy states form
a manifold of degenerate field states disjoint from the primitive vacuum
state; in the present case, we
assume that the bare masses are positive, and that the higher-order
interaction energy terms give rise to physical vacuum states
having complex energy eigenvalues, i.e., are closed channels.)
In order to gain physical insight concerning this possibility,
we devote the remainder of this
section and of the paper to a nonperturbative calculation
on vacuum states and energies.  With minor modifications
the mathematics that follows could accomodate 
the vacuum state matrix of any suitable Hamiltonian;
to keep to a specific and simple model, we use the Hamiltonian
of Eq.\ \eqref{E6:eq10}.
We establish a
two-channel problem consisting of the vacuum states
\begin{equation}\label{E7:eq1}
\Psi^F\ =\ \left[\begin{matrix} \Upsilon(F,0)\\
0\end{matrix}\right],\qquad
\Psi^B\ =\ \left[\begin{matrix} 0\\
\Upsilon(B,0)\end{matrix}\right].
\end{equation}
In the time interval $[0,\tau]$, let the  
normalized input state be (cf. Eq.\ \eqref{E4:eq20})
\begin{equation}\label{E7:eq2}
\Phi_{\text{in}}(0,\tau)\ =\  
\left[\begin{matrix} \Upsilon(F,0)\cos\theta\\
\Upsilon(B,0)\exp(i\psi)\sin\theta\end{matrix}\right],
\end{equation}
where $\theta$ and $\psi$ are polar and
azimuthal coordinates, respectively, on $S^2$.
The output state also comprises the direct sum of
vacuum states taken at two different times (cf. Eq.\ \eqref{E4:eq21}):
\begin{equation}\label{E7:eq3}
\Phi_{\text{out}}(\tau,0)\ =\  
\left[\begin{matrix} \Upsilon(F,0)\beta^F\\
\Upsilon(B,0)\beta^B\end{matrix}\right],
\end{equation}
where $\beta^F$ and $\beta^B$ are complex
coefficients that comprise the output data,
which we know beforehand must satisfy the normalization condition
\begin{equation}\label{E7:eq4}
|\beta^F|^2\,+\,|\beta^B|^2\ =\ 1.
\end{equation}
We assume a
time-dependent state vector $\Psi(t)$ of the form
\begin{equation}\label{E7:eq5}
\Psi(t)\ =\ \Psi^F\,\Phi_F(t)\,+\ \Psi^B\,\Phi_B(t),
\end{equation}
and establish a coupled, first-order 
differential equation for the time evolution
of the coefficient functions $\Phi_Y(t)$, $Y=F,B$.
The equations of motion are
\begin{equation}\label{E7:eq6}
i\frac{d}{dt}\Phi_Y(t)\ =\ \alpha_Y \sum_{Y'=F,B}
(\Psi^Y;H\Psi^{Y'})
\Phi_{Y'}(t).
\end{equation}
The matrix of the Hamiltonian proves to be
\begin{equation}\label{E7:eq7}
(\Psi^Y;H\Psi^{Y'})\ =\ E^{[0]}\alpha_Y\delta^{YY'}
\,+\,E^{[1]}\bigl(\alpha_Y\zeta^Y \delta^{YY'}
-\xi\delta^{YF}\delta^{BY'}
-\xi\delta^{YB}\delta^{FY'}\bigr).
\end{equation}
In Eq.\ \eqref{E7:eq7}, $E^{[0]}$ and $E^{[1]}$ 
are the conventional (FMT only) vacuum expectation values of the
zero-order Hamiltonian and $(1/4)\int\phi(x)^4 d^3x$,
respectively; to be sure, both of these quantities are plus
infinity in the present theory, but we shall pretend otherwise
and see what happens.  The eigenvalues of the Hamiltonian
matrix are
\begin{equation}\label{E7:eq8}
E^Y\ =\ E^{[0]}\,+\,E^{[1]}\bigl[(1/2)\bigl(\zeta^F+\zeta^B\bigr)
-i\alpha_Y\sqrt{-D}\bigr],
\end{equation}
where $\alpha_Y$ is defined in Eq.\ \eqref{E5:eq9.5},
and we have presumed that the $D$ of Eq.\ \eqref{E6:eq20}
is negative.  We therefore have a coupled-channel
problem that is akin to an ordinary
single-channel 
bound state problem in the context of a second-order, time-independent
Schr\"{o}dinger equation;
however, there is no energy-like parameter that can be varied here,
nor is there a segment of the time axis in which a
shift between a rising and falling exponential
can occur, hence a bound state in the time
dimension does not occur in this case.
 
     Continuing the argument, we define
\begin{subequations}\label{E7:eq9}
\begin{align}
\bar E\ &=\ E^{[0]}\,+\,E^{[1]}(\zeta^F +\zeta^B)/2,
\label{E7:eq9a}\\
\kappa\ &=\ (\zeta^F-\zeta^B)/2, \label{E7:eq9b}\\
\mu\ &=\ \sqrt{\xi^2-\kappa^2}\ =\ \sqrt{-D}\ >\ 0,\label{E7:eq9c}\\
\cos\sigma\ &=\ \kappa/\xi,\label{E7:eq9d}\\
\sin\sigma\ &=\ \mu/\xi.\label{E7:eq9e}
\end{align}
\end{subequations}
A set of eigensolutions to the Schr\"odinger equation 
Eq.\ \eqref{E7:eq6} is then, for $Y=F,B$,
\begin{equation}\label{E7:eq10}
\Phi^{(Y)}(t)\ =\ \left[\begin{matrix}
\Phi^{(Y)}_F(t)\\ \Phi^{(Y)}_B(t)
\end{matrix}\right]\ =\ 
\exp(-i\bar E t -\alpha_Y \mu E^{[1]}t)
\left[\begin{matrix} i\xi \\ i\kappa-\alpha_Y\mu
\end{matrix}\right]
\end{equation}
The reason for the superscript is that the
solution $\Phi^{(F)}(t)\  (\Phi^{(B)}(t))$ 
decreases exponentially
as $t\to+\infty\ (t\to-\infty)$. 
The matrices of $\eta$ and $\eta H$ in the latter basis are
time-independent, and have the values
\begin{subequations}\label{E7:eq11}
\begin{align}
\Phi^{(Y)}(t)^\dagger\eta\Phi^{(Y')}(t)
\ &=\ \delta^{YF}\delta^{BY'}[2\mu(i\kappa+\mu)]
+\delta^{YB}\delta^{FY'}[-2\mu(i\kappa-\mu)],
\label{E7:eq11a}\\
\Phi^{(Y)}(t)^\dagger\eta H\Phi^{(Y')}(t)
\ &=\ \delta^{YF}\delta^{BY'}[2\mu(i\kappa+\mu)
(\bar E+i\mu E^{[1]})]\nonumber\\
&\ \ \ +\delta^{YB}\delta^{FY'}[-2\mu(i\kappa-\mu)
(\bar E-i\mu E^{[1]})].\label{E7:eq11b}
\end{align}
\end{subequations}
A general solution to the Schr\"odinger equation has the form
\begin{equation}\label{E7:eq12}
\Phi(t)\ =\ C^{(F)}\Phi^{(F)}(t)
\,+\,C^{(B)}\Phi^{(B)}(t).
\end{equation}
We choose the constants $C^{(Y)}$ so that the 
input boundary conditions
Eq.\ \eqref{E7:eq2} are satisfied.  We find that
\begin{multline}\label{E7:eq13}
C^{(Y)}\ =\ 
\bigl[-\alpha_Y\sin\theta\exp(i\psi+i\bar E\tau)
+\alpha_Y\cos\theta\exp(-i\alpha_Y\sigma+\alpha_Y\mu E^{[1]}\tau)
\bigr]\\
\ \ \ \ \ \times\bigl\{i\xi\bigl[\exp(-i\sigma+\mu E^{[1]}\tau)
-\exp(+i\sigma-\mu E^{[1]}\tau)\bigr]\bigr\}^{-1}.
\end{multline}
The expectation value---as defined in Eq.\ \eqref{E4:eq25}---for
the unit operator and the Hamiltonian in the state $\Phi(t)$
are given by
\begin{subequations}\label{E7:eq14}
\begin{align}
[I]_{\rm{Av}}\ &=\ 2\sin\sigma\bigl[-\sin\sigma
+2\sin\sigma\cos(\sigma+\psi+\bar E\tau)\cosh(\mu E^{[1]}\tau)
\nonumber\\
&\ \ \ -2\cos\sigma\sin(\sigma+\psi+\bar E\tau)\sinh(\mu E^{[1]}\tau)\bigr]
\Delta^{-1},\label{E7:eq14a}\\
[H]_{\rm{Av}}\ &=\ \bar E [I]_{\rm{Av}}\,+\, E^{[1]}A\Delta^{-1},
\label{E7:E14b}\\
\Delta\ &=\ \cosh(2\mu E^{[1]}\tau)
\,-\,\cos(2\sigma),\label{E7:eq14c}\\
A\ &=\ 2\xi\sin^2\sigma\bigl[\cos\sigma-2\cos\sigma
\cos(\sigma+\psi+\bar E\tau)\cosh(\mu E^{[1]}\tau)\nonumber \\
&\qquad -2\sin\sigma\sin(\sigma+\psi+\bar E\tau)\sinh(\mu E^{1]}\tau)
\bigr].\label{E7:eq14d}
\end{align}
\end{subequations}
When $\mu E^{[1]}\tau$ is large, we find that
\begin{subequations}\label{E7:eq15}
\begin{align}
[I]_{\rm{Av}}\ &\to\ -4\sin{\sigma}\sin(\psi+\bar E\tau)
\exp(-\mu E^{[1]}\tau)\,+\,O\bigl(\exp(-2\mu E^{[1]}\tau)\bigr),
\label{E7:eq15a}\\
[H]_{\rm{Av}}\ &\to\ -4\sin\sigma\bigl[\bar E\sin(\psi+E^{1]}\tau)
\nonumber\\
&+E^{[1]}\xi\sin\sigma\cos(\psi+\bar E\tau)\bigr]
\exp(-\mu E^{[1]}\tau)
\,+\,O\bigl(\exp(-2\mu E^{[1]}\tau)\bigr).\label{E7:eq15b}
\end{align}\end{subequations}
Hence, if the ``experiment'' is performed over a time
interval $\tau$ that is sufficiently long, the vacuum expectation values
across a time=constant surface of the
probability and of the energy are both
exponentially small uniformly over the closed time
interval $[0,\tau]$.   In other words, whatever be the input
vacuum state, the magnitude and phase of the resulting
time-dependent vacuum state will, for a sufficiently
long time $\tau$, adjust itself so that, at any given time, 
almost equal amounts of probability 
are in FMT and in BMT, and almost equal amounts of energy are 
in FMT and in BMT.  

     Suppose, finally, that we compute the
expectation values $\bar T^{\mu\nu}(x^0,\mathbf{x})$
with respect to $\Phi(x^0)$
of the components of the stress-momentum-energy-density operators
$T^{\mu\nu}(\mathbf{x})$, $\mu,\nu=0,1,2,3$, as given
in Eqs.\ \eqref{E6:eq8}, \eqref{E6:eq12} and \eqref{E6:eq12x1}:
\begin{equation}\label{E7:eq16}
\bar T^{\mu\nu}(x^0,\mathbf{x})\ =\ \Phi(x^0)^\dagger\,\eta\,
T^{\mu\nu}(\mathbf{x})\,\Phi(x^0).
\end{equation}
So long as $\Phi(x^0)$ satisfies the Schr\"odinger equation
Eq.\ \eqref{E6:eq14}, the position-dependent array 
$\bar T^{\mu\nu}(x^0,\mathbf{x})$ can be shown to have
zero four-divergence and to have the transformation
properties of a second-rank contravariant tensor field
under the action of the restricted Poincar\'e group, in the
sense that the application of one of the Lie algebra
elements of Eq.\ \eqref{E6:eq16} to $\Phi(x^0)$ yields the
same effect on $\bar T^{\mu\nu}(x^0,\mathbf{x})$
as would have the corresponding Lie algebra element 
acting on such a tensor field.  Accordingly, we can take
such a $\bar T^{\mu\nu}(x^0,\mathbf{x})$ to be
the source distribution of a linearized, classical 
gravitational field in a background Minkowski space-time.
If we choose $\Phi(x^0)$ to be the vacuum state of 
Eqs.\ \eqref{E7:eq5}, \eqref{E7:eq12} and \eqref{E7:eq13}, 
the result of Eq.\ \eqref{E7:eq16}
is not a tensor field (in particular, with respect to Lorentz boosts),
since the vacuum state
does not satisfy the complete Schr\"odinger equation.
Nevertheless, we take the (still divergent)
vacuum expectation value $\bar T^{00}(x^0,\mathbf{x})$ 
to be an estimate for the energy density due to the vacuum.
This energy density amounts to the expectation value for
total energy, divided by the total volume of space.
We construe the result Eq.\ \eqref{E7:eq15a}
as contributing to an
explanation for the cosmological constant problem,
as described in, say, Weinberg \cite{R:Weinberg2} or Carroll  
\cite{R:Carroll1}:
Given that the discriminant $D$ of Eq.\ \eqref{E6:eq20} is negative, 
the net vacuum energy density in space-time should
have a very small magnitude, and the
energy density would depart from zero due mainly to the presence of ordinary
matter in FMT or BMT, and possibly to small vacuum effects that
do not enter into the present simple theory and approximation.
Eq.\ \eqref{E7:eq15} also suggests that if $D<0$ the probability
that the system is found to be in the BMT vacuum
state, but not necessarily in states involving matter in BMT,
is about the same as the probability of finding the FMT vacuum state.

     \{Remarks added in the arXiv version:  
Given that the above-described ``antiparallel''
world exists, a possible circumstance is that there is more
matter in BMT states than in FMT states.  
Accordingly, the net average energy density
in the cosmos---the $\bar T^{00}(x^0,\mathbf{x})$ of \eqref{E7:eq16}---would 
be negative on large distance scales, and a modification of Einstein's
field equations for gravitation would not be needed to explain
the phenomenon of so-called dark energy, which is described in, say, articles
cited at the web location\newline 
http://supernova.lbl.gov/\~{ }evlinder/sci.html .\}
 
\section{Appendix:  Transition rates} \label{S:App}

     We want to obtain an expression that permits us
to deal with the energy delta-functions in Eq.\ \eqref{E5:eq20}
to obtain transition probabilities per unit time and cross sections.
Although the formalism permits inputs at the initial
and final times to be coherent, and permits the study of outputs
with definite phase relationships between the temporally
earlier and temporally later parts of the output, we shall
not attempt this level of generality here:  we assume
phase incoherence between the FMT and the BMT parts of
the input, and discard all information on interference
between the FMT and the BMT parts of the output.
In other words, we shall presume a block-diagonal (FF and BB only) density
matrix at input, and discard block off-diagonal (FB and BF) parts of
the density matrix at output. 

     Let us begin with Eq.\ \eqref{E5:eq18} with $\gamma'\neq\gamma$,
with the adiabatic switching factors $\exp[-\epsilon|t-t_1|]$
inserted in the integrands, and the integrals carried out:
\begin{equation}\label{App:eq1}
\begin{split}
\bigl(\Phi&^{[0]R,Y'}_{E'\gamma'}(t);\Phi^{R,Y}_{E\gamma}(t)\bigr)
\ =\ -i\bigl(\Psi^{[0]R,Y'}_{E'\gamma'};\eta T(E)\Psi^{[0]R,Y}_{E\gamma}
\bigr)\\
&\times\Bigl[\delta^{Y'F}
\frac{i\exp[i(E'-E)t]}{(E-E'+i\epsilon)}
\,+\,\delta^{Y'B}
\frac{i\exp[i(E'-E)t]}{(E-E'-i\epsilon)}\Bigr].
\end{split}
\end{equation}
We relate the parameter $\epsilon$ to the effective on time $\tau$
of the interaction as follows: insofar as the interaction affects the 
(say) FMT output, we have presumed that the 
FMT part of the Green's function $G^{[0]}(t-t_1)$
is switched on as $\exp[\epsilon(t_1-t)]$ and therefore has squared
magnitude $\exp[2\epsilon(t_1-t)]$.  We have
\begin{equation}\label{App:eq2}
\tau\ =\ \int_{-\infty}^t \exp[2\epsilon(t_1-t)]dt_1\ =\ 1/(2\epsilon).
\end{equation}
To obtain a transition probability
per unit time, we shall divide the transition probability,
summed over a range in energy of output states, by $\tau$.  A similar result
obtains for the effect of the modulated Green's function
on the BMT output.

     We compute the absolute square of either the FMT ($Y'=F$)
or the BMT ($Y'=B$) part of the rhs of Eq \eqref{App:eq1}.
In both cases, the rhs has a factor $1/[(E-E')^2+\epsilon^2]$.
This factor will be construed as tending to a delta-function in energy
as $\epsilon$ becomes small, in fact close to
$(\pi/\epsilon)\delta(E-E')$.  Since $\pi/\epsilon=2\pi\tau$,
the transition probability per unit time becomes, when the sum
over output energy states is converted to an integral with a density
of states,
\begin{equation}\label{App:eq3}
\frac{2\pi}{\hbar}|
\bigl(\Psi^{[0]R,Y'}_{E\gamma'};\eta T(E)\Psi^{[0]R,Y}_{E\gamma}
\bigr)|^2 \rho^{Y'}_{\gamma'}(E),
\end{equation}
where $\rho^{Y'}_{\gamma'}(E)$ is a density in energy of output states
of type $Y'=F$ or $Y'=B$, state index $\gamma'$, and energy $E$.

\pagebreak

\end{document}